\documentclass[12pt]{iopart}
\usepackage{epsfig}
\usepackage{subfigure}

\usepackage{iopams}  

\def\ave#1{\langle {#1} \rangle}

\begin{document}

\title[Event-by-event fluctuations and the QGP phase transition]{
Event-by-event fluctuations and the QGP phase transition}

\author{Tapan K. Nayak \footnote{On leave from Variable Energy Cyclotron Centre, Kolkata, India}}
\address{CERN, CH-1211, Geneva-23, Switzerland}
\ead{Tapan.Nayak@cern.ch}

\begin{abstract}
Fluctuations of thermodynamic quantities are fundamental 
to the study of QGP phase transition. Event-by-event fluctuations
of many quantities have been studied by dedicated heavy-ion experiments.
A brief review of recent
experimental results is presented. The prospect
for future study of fluctuations is discussed.
\end{abstract}


\section{Introduction}

Fluctuations are of fundamental importance for studying perturbation of a 
thermodynamic system. Several 
thermodynamic quantities show varying fluctuation patterns when the system undergoes
a phase transition. Moreover, fluctuations are sensitive to the nature of the
transition. Of particular interest to us is the fluctuation induced by the 
phase transition between normal hadronic matter and the quark-gluon plasma (QGP), 
a condition similar to that existed within a few tens of microseconds of the early
universe. Event-by-event fluctuations of thermodynamic quantities measured 
in high energy heavy-ion collisions provide a reasonable framework
for studying the nature of the QGP phase transition in the laboratory \cite{EBYE:rev}.
Large fluctuations in energy density are expected if the phase transition is
of first order whereas a second order phase transition may lead to a
divergence in specific heat. An increase in the fluctuation in energy 
density for a second order transition 
may be expected due to long range correlations in the system. 
Furthermore, the prospect of locating the critical point of the QGP phase 
transition where the fluctuations are predicted to be largely enhanced \cite{EBYE:tricrit}
makes this study rather interesting and challenging.

The rapid development in the field of event-by-event fluctuations in
recent years is related to the availability of large amount of high
multiplicity data from 
dedicated heavy-ion experiments at the CERN-SPS and BNL-RHIC. 
Although tremendous progress has been made in our understanding of 
the fluctuations, still the promise of the fluctuation measures to 
provide a clear understanding of the QGP phase transition has yet to be realized.
Recent advances in theoretical aspects, especially from
lattice computations, and plans for dedicated experiments give much
needed assurance.
There are several effects which make the study of fluctuations non-trivial.
One of the most important issues concerns the separation of statistical 
and known physics contributions from the measured fluctuations in order to
identify the dynamical part associated with the phase transition.

In case of high energy heavy-ion 
collisions, there are several sources which contribute to the measured fluctuation,
which include: (a) geometrical (impact parameter, number of participants, 
detector acceptance), (b) energy, momentum and charge conservation,
(c) anisotropic flow, (d) Bose-Einstein correlations,
(d) resonance decays, (e) contribution from jets and mini-jets, etc.
These effects must be taken into account in order to infer
about the fluctuation from dynamical origin related to that of the phase transition. 
Present analyses use one or more of the following
methods for this purpose:
\begin{itemize}
\item construction of mixed events which take care of instrumental effects, 
\item simulation of fluctuations originating from statistical and known 
      physics sources,
\item construction of fluctuation measures which are robust against 
      known fluctuations.
\end{itemize}
We give a brief review of some of the fluctuation measures which have been 
studied in heavy-ion collisions. These include fluctuations 
in multiplicity, particle ratio, net charge, mean transverse momentum $\ave{p_{\rm T}}$, and
methods of balance functions and long range correlations.
Finally we give an outlook on the
near term plans including the prospect of studying event-by-event physics
in the ALICE experiment at the LHC.

 \section{Multiplicity fluctuations} 
\begin{figure} 
  \begin{center}
    \leavevmode
    \epsfig{file=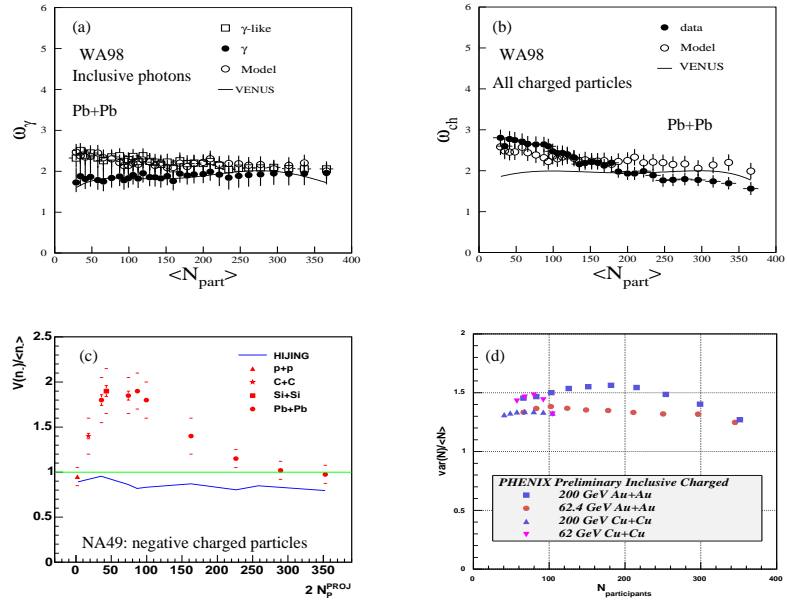, width=4.3in}
  \end{center}
\vspace*{-1cm}
  \caption{Multiplicity fluctuations at SPS and RHIC energies.
}
  \label{mult_fluc}
\end{figure}

Multiplicity of produced particles characterizes the evolving system in a 
heavy-ion collision and thus fluctuation in multiplicity may provide a distinct
signal of the QGP phase transition \cite{EBYE:rev,EBYE:wa98_ebye}. 
Since multiplicity distributions for narrow centrality bins can be described by Gaussian 
distributions, their fluctuations are expressed in terms of scaled variance, 
defined as, $\omega=var(N)/\ave{N}$, where $\ave{N}$ and $var(N)$ represent the variance and
mean of the multiplicity distribution, respectively.
Figure~\ref{mult_fluc} shows observed scaled variance for SPS and RHIC 
energies. The results from the WA98 experiment \cite{EBYE:wa98_ebye}, those corresponding to
photons and charged particles, are compared to different model calculations.
Although the 
experimental data is consistent with the model calculations within the
quoted error bars, the increasing trend of fluctuation for charged particles
towards peripheral collisions is clearly visible. 
The scaled variance of charged particles as a function of centrality
from NA49 \cite{EBYE:na49_mult_fluc} shows a non-monotonic behaviour. 
The PHENIX data \cite{EBYE:phenix_mult_fluc} for Cu--Cu collisions 
at $\sqrt{s_{\rm NN}} = 62.4$~GeV shows
a small structure for non-central collisions whereas at higher energies the data are smoother.
Detailed understanding of these results would require considerations of
centrality selection and detector effects.

 \section{Net charge fluctuations}
\begin{figure} 
  \begin{center}
    \leavevmode
    \epsfig{file=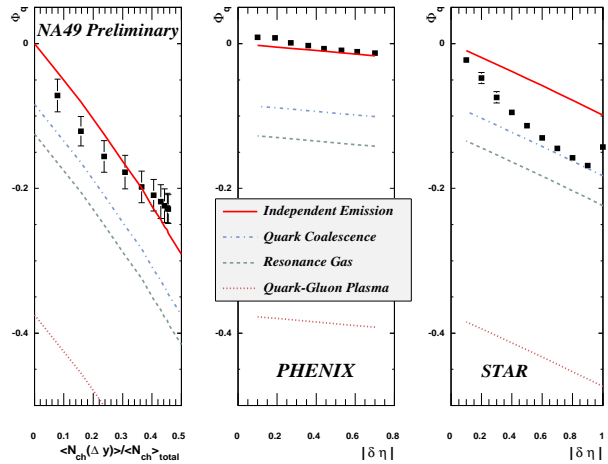, width=3.4in,height=3.1in}
  \end{center}
\vspace*{-1cm}
  \caption{Dynamical fluctuation of net charge for NA49, PHENIX and STAR 
  experiments \cite{EBYE:mitchell}.
}
  \label{netcharge_fluc}
\end{figure}

Fluctuations of conserved quantities like electric charge, baryon number or
strangeness are predicted to be significantly reduced in a QGP scenario
as they are generated in the early plasma stage of the system created in heavy-ion 
collisions with quark and gluon degrees of freedom \cite{EBYE:As00,EBYE:Ko00}.
The fluctuation generated at the QGP stage will increase 
as the system evolves in time \cite{EBYE:Sh01,EBYE:evolution}.
Net charge fluctuations have been measured by experiments at SPS and RHIC
using different fluctuation measures. 
Among these are $\Phi_q$ of NA49 \cite{EBYE:na49_net},
$\nu_{+-,dyn}$ of STAR \cite{EBYE:star_mult} and $v(Q)$ as well as $\nu_{+-,dyn}$ 
used by PHENIX \cite{EBYE:phenix_net}. A common framework 
which relates these variables \cite{EBYE:volo} has been used to compile
the available results \cite{EBYE:mitchell}. This is shown in Fig.~\ref{netcharge_fluc},
along with predictions from independent particle emission, quark coalescence, resonance
gas and a QGP scenario. Both NA49 and PHENIX results are consistent with the 
independent particle emission scenario, whereas the result for 
STAR is close to the case of the quark coalescence model. 

\begin{figure} 
  \begin{center}
    \leavevmode
    \epsfig{file=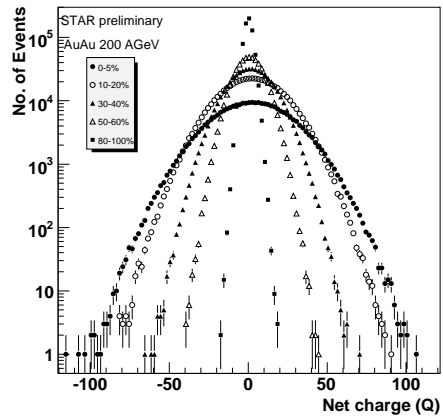, width=2.5in}
  \end{center}
\vspace*{-0.5cm}
  \caption{Net charge distributions of particles with $p_{\rm t}$ below 1GeV/c
for Au--Au collisions at $\sqrt{s_{\rm NN}} = 200$~GeV
at different centralities. 
}
  \label{netcharge_moments}
\end{figure}

Recently, lattice computations \cite{EBYE:ekr,EBYE:ggnls} have been performed
to study hadronic fluctuations. These calculations predict an enhancement of
fluctuation in the hadronic phase and suppression of fluctuations in the high
temperature phase of the QGP. More interestingly, prominent structure in the
higher order moments of net charge distributions 
have been observed for temperatures close to the transition
temperature. The higher moments of net charge distributions can be studied
from experimental data. Figure~\ref{netcharge_moments} shows the 
net charge distributions of particles
with $p_{\rm T}$ below 1GeV/c for Au--Au collisions at $\sqrt{s_{\rm NN}} = 200$~GeV for 
different centralities in the STAR experiment within a pseudorapidity coverage
of $-1\le \eta \le 1$. Efforts are underway to study higher order moments of
these distributions by making smaller bins in detector acceptances and $p_{\rm T}$. 

 \section{Fluctuations of particle ratio}
Relative production of different types of particles produced in the hot and dense matter
might be affected once the system goes through a phase transition.
Of particular interest is the strangeness
fluctuation in terms of the ratio of kaons to pions. 
Large broadening in the yields of kaons to pions
has long been predicted because of the differences in free enthalpy of the
hadronic and QGP phase. This could be probed through the fluctuation in
the $K/\pi$ ratio.

\begin{figure}
\hspace{3pc}%
\begin{minipage}{16pc}
\includegraphics[width=16pc]{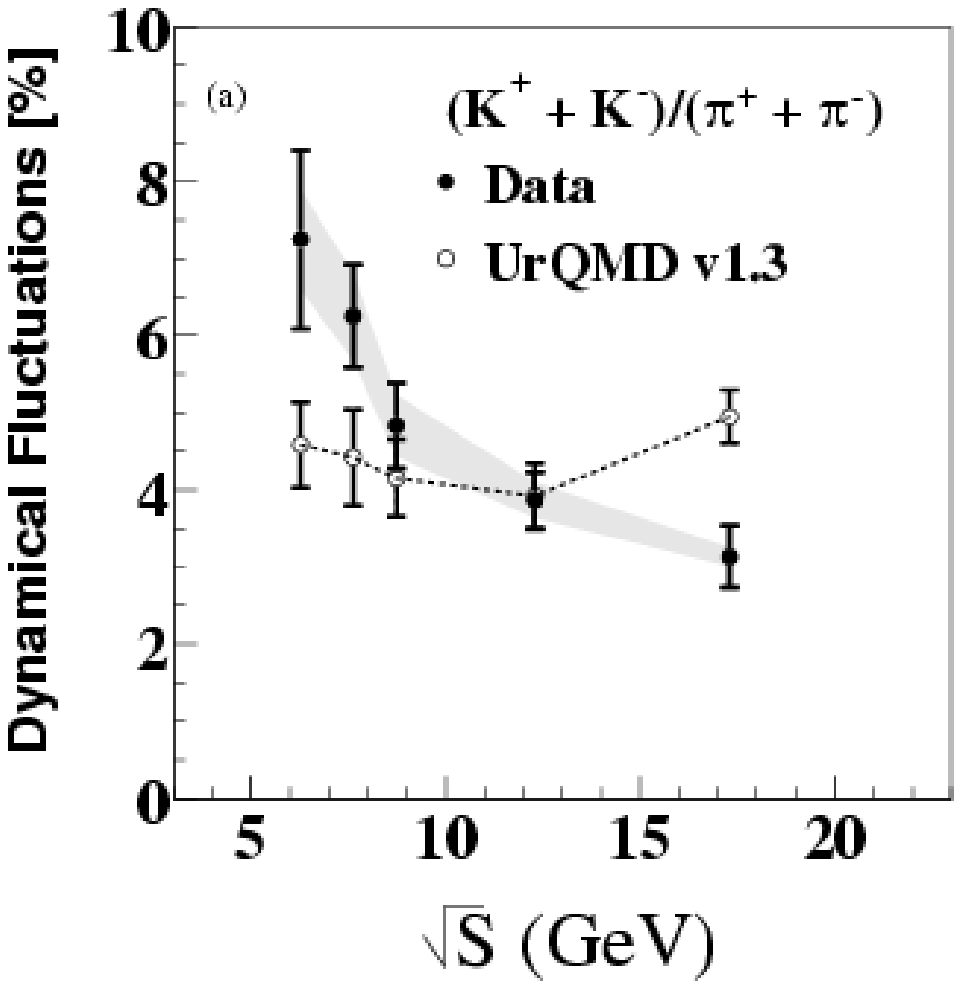}
\end{minipage}\hspace{0.1pc}%
\begin{minipage}{14pc}
\includegraphics[width=14pc]{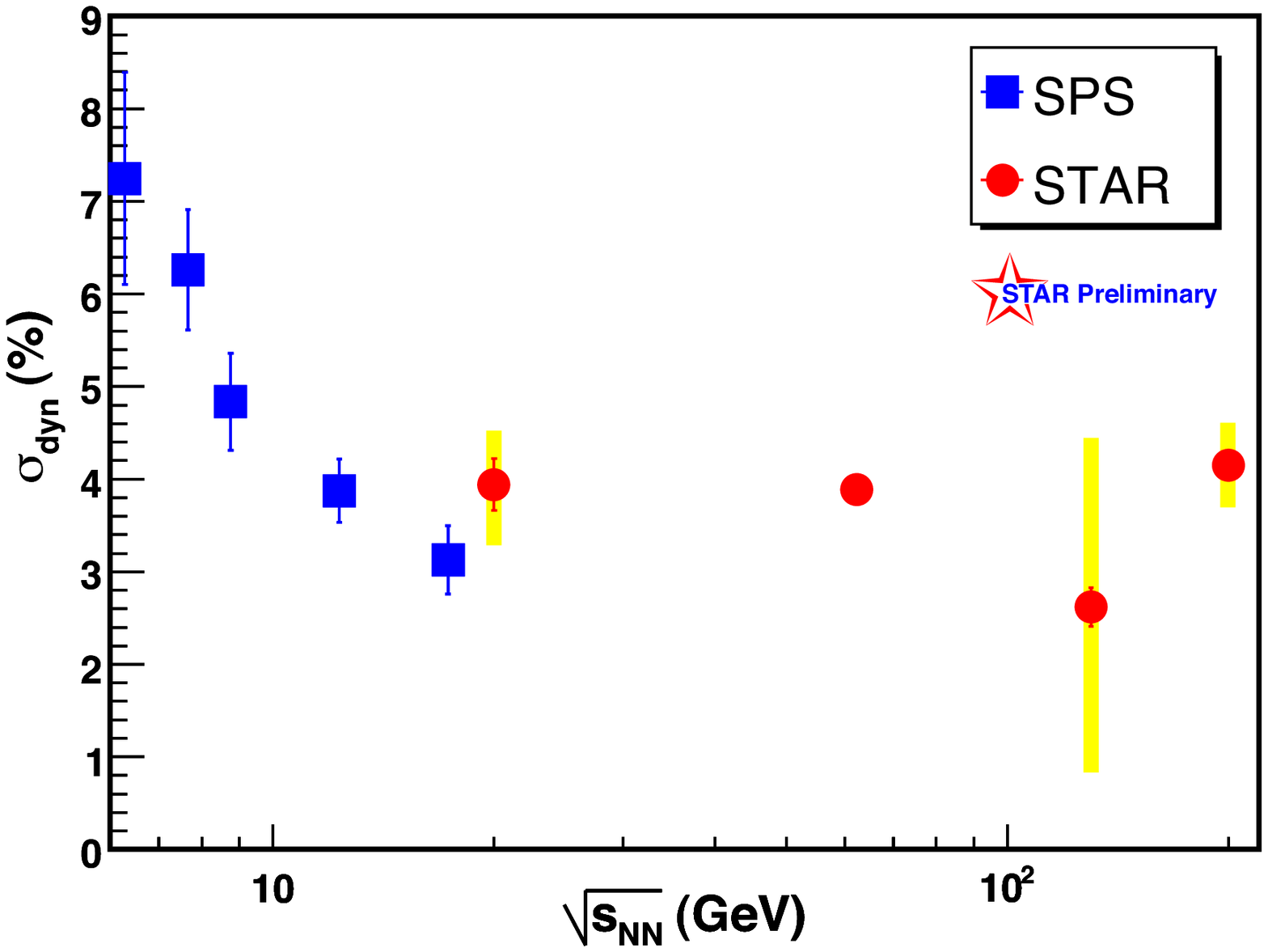}
\end{minipage} 
\caption{Excitation function for $\sigma_{dyn}$ of 
$[K^{+}+K^{-}]/[\pi^{+}+\pi^{-}]$ ratio at the SPS (left panel) and
with an extension to RHIC (right panel).}
\label{edep}
\end{figure}

A detailed study at SPS has been carried out at several beam energies \cite{EBYE:qm04k2pi}. 
The ratio of inclusive mid-rapidity yields of
$\ave{K^-}/\ave{\pi^->}$ has an increasing trend with beam energy, whereas a
horn structure is seen in the ratio of  $\ave{K^+}/\ave{\pi^+}$. It has been shown that 
the dynamical fluctuations ($\sigma_{\rm dyn}$)
in the ratio of $p/\pi$ has an increasing trend with 
respect to beam energy which could be explained by model calculations.
At the same time $\sigma_{\rm dyn}$ in the $K/\pi$ ratio
is seen to decrease with beam energy, a behavior which could not be explained by the
same model. The $\sigma_{\rm dyn}$ values at SPS energies are shown in the left panel
of Fig.~\ref{edep}. 
The STAR experiment has performed a similar study on the 
event-wise fluctuations of 
the $K/\pi$ ratio for \mbox{Au--Au} collisions at $\sqrt{s_{\rm NN}} = 62.4$~GeV and 
$\sqrt{s_{\rm NN}}=200$~GeV \cite{EBYE:star_kpi}. 
A reduction as a function of
centrality is reported for the two energies. The right panel of Fig.~\ref{edep}
shows an excitation energy plot for $K/\pi$ ratio
extended up to the highest RHIC energies.
The fluctuation decreases with increasing energy up to the highest SPS energy and 
remains constant at higher RHIC energies. Theoretical investigations
\cite{EBYE:torr} are underway to explain such behaviour.

\section{$\ave{p_{\rm T}}$ fluctuations}
The $\ave{p_{\rm T}}$ of emitted particles in an event is related
to the temperature of the system. Thus the event-by-event fluctuations of
average $p_{\rm T}$ is sensitive to the temperature fluctuations predicted for 
the QGP phase transition. $\ave{p_{\rm T}}$ can be measured experimentally
with high precision. The interpretation of the the results has to include
considerations of acceptance effects, volume fluctuations, resonance decays, 
elliptic flow, HBT correlations, hard scattering and jet production. 
Several measures of fluctuation
have been introduced in order to probe the dynamical fluctuation from the
measured values.
 \begin{figure} 
  \begin{center}
    \leavevmode
    \epsfig{file=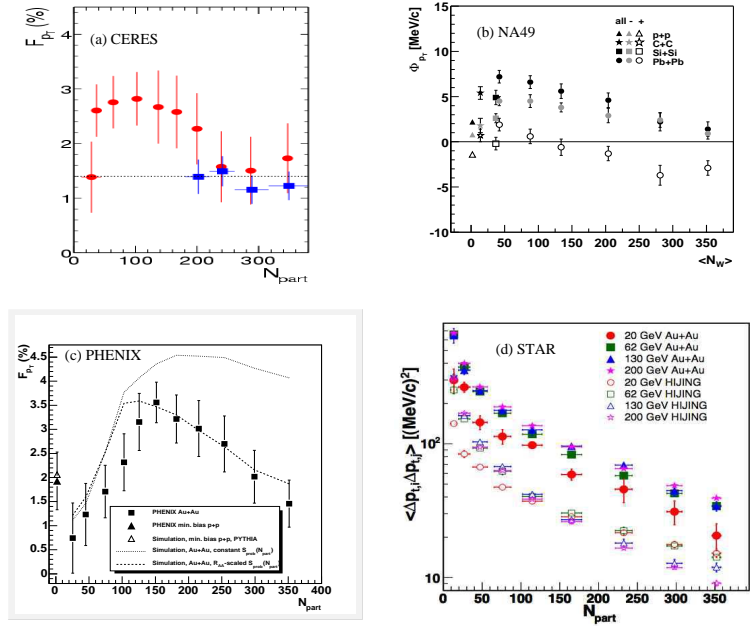,width=4in}
  \end{center}
\vspace*{-0.5cm}
  \caption{Dynamical $\ave{p_{\rm T}}$ fluctuations as a function of
centrality of the collision from (a) CERES, (b) NA49, 
(c) PHENIX and (d) STAR experiments.
}
  \label{pt_fluc}
\end{figure}
Figure~\ref{pt_fluc} shows 
the centrality dependence of dynamical fluctuations reported by
CERES \cite{EBYE:ceres_pt}, NA49 {\cite{EBYE:na49_pt}, 
PHENIX \cite{EBYE:phenix_pt} and STAR \cite{EBYE:star_pt}. The results
presented in Fig.~\ref{pt_fluc}(d) show a smooth variation of
fluctuation with centrality whereas the other measurements show
non-monotonic behaviour. Efforts are being made to understand
the nature and origin of these fluctuations. Because of the choice
of several variables, extraction of an excitation energy plot
combining data from SPS to RHIC is not straightforward. It is of
interest to us to have a common framework for presenting
the results from different experiments. In order to be
more sensitive to the origin of fluctuations, differential
measures have been adopted where the analysis is performed at
different scales (varying bins
in $\eta$ and $\phi$).
The scale dependence of $\ave{p_{\rm T}}$ fluctuation for
three centralities in \mbox{Au--Au} collisions at
$\sqrt{s_{\rm NN}}=200$~GeV  \cite{EBYE:star_pt_etaphi} is shown 
in Fig.~\ref{pt_etaphi}. The extracted autocorrelations are
seen to vary rapidly with collision centrality, suggesting that
fragmentation is strongly modified by a dissipative medium in
more central collisions relative to peripheral collisions. Further
studies for different charge combinations will provide more
detailed information.
 \begin{figure}
  \begin{center}
    \leavevmode
    \epsfig{file=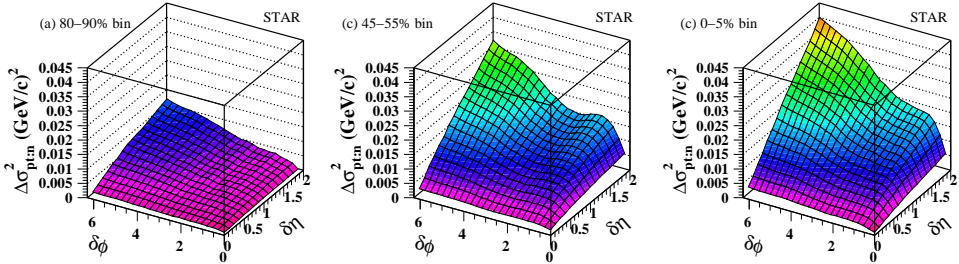,width=5in}
  \end{center}
\vspace*{-0.5cm}
  \caption{Scale dependence of $\ave{p_{\rm T}}$ fluctuation within
the STAR acceptance expressed in terms of per-particle variance difference
as discussed in \cite{EBYE:star_pt_etaphi}.}
  \label{pt_etaphi}
\end{figure}

 \section{Balance functions}
The method of Balance Functions (BF) \cite{EBYE:Pratt},
provides a measure of correlation of oppositely charged particles produced during
heavy-ion collisions. The basic idea is that the charged hadrons are produced locally
as oppositely charged-particle pairs. The particles of such a pair are separated
in rapidity due to the initial momentum difference and secondary interaction with other
particles. The particles of a pair produced earlier are separated further in rapidity
compared to the particles coming from a pair produced later in time. Since the width of the 
correlation can be related to the time of hadronization of the charged particles,
this would signal any possible delayed hadronization, corresponding to the formation of 
a high density QGP matter.

Both STAR \cite{EBYE:star_bf} and NA49 \cite{EBYE:na49_bf} experiments have
made detailed measurements of the BFs for
various colliding systems, centralities,
pseudorapidity intervals as well as for identified charged particles.
Here we report two of these studies; centrality dependence 
and excitation energy dependence of BF widths.
The left panel of Fig.~\ref{bf} shows the width of the BFs as 
function of the normalized impact parameter for \mbox{Pb--Pb} collisions 
at $\sqrt{s_{\rm NN}}$=17.2~GeV and \mbox{Au--Au} collisions at 
$\sqrt{s_{\rm NN}}$=130~GeV.
The widths of the BF decrease from peripheral to central collisions 
in experimental data whereas the shuffled data shows no such reduction.
The decrease in the width can be quantified by the use of a normalized parameter, 
$W$, expressed as enhancement in the width in the data with respect to the
corresponding shuffled values.
The values of $W$ are plotted in the right panel of Fig.~\ref{bf} as a function of
beam energy \cite{EBYE:na49panos}. The increase of the $W$
from SPS to RHIC may be interpreted in terms of a delayed hadron scenario.
\begin{figure}
\hspace{5pc}%
\begin{minipage}{12pc}
\includegraphics[width=12pc]{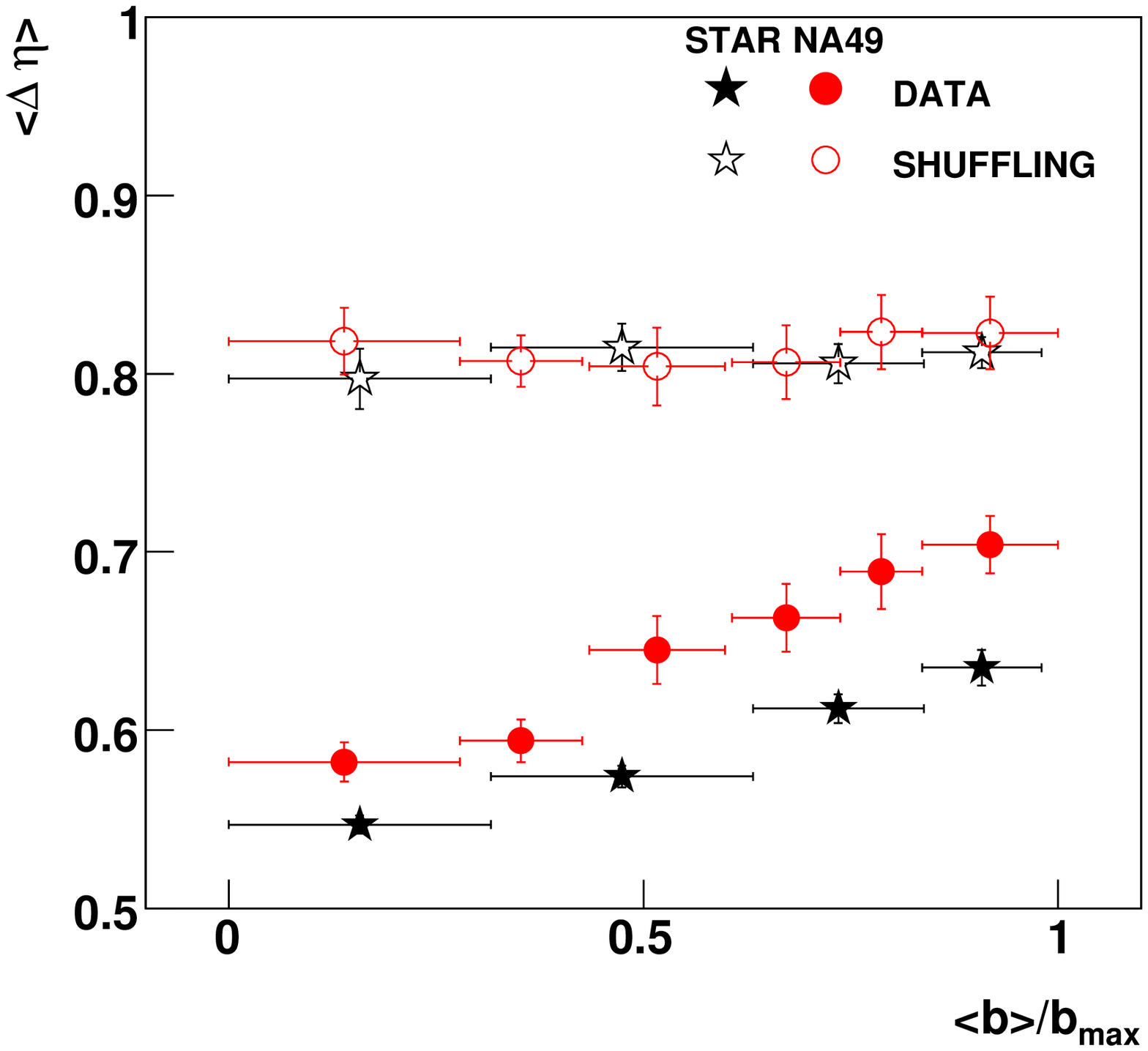}
\end{minipage}\hspace{0.1pc}%
\begin{minipage}{12pc}
\includegraphics[width=12pc]{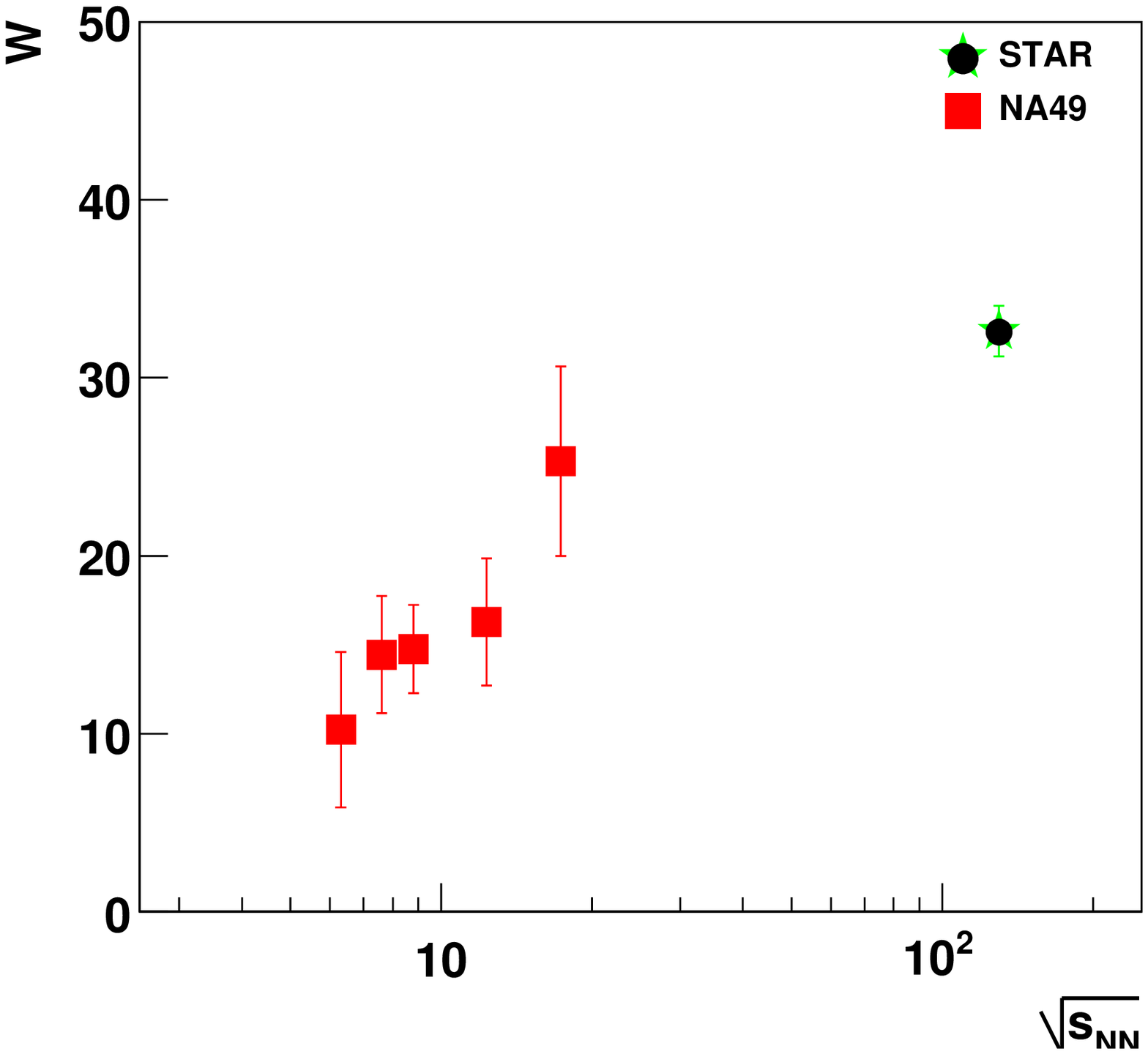}
\end{minipage} 
\caption{(a) The width of the balance function as a function of centrality for
experimental data along with results for shuffled bins (b) Normalized parameter ($W$) of
balance function as a function of beam energy.
}
\label{bf}
\end{figure}

 \section{Summary and outlook}
Experiments at SPS and RHIC have given a wealth of data on fluctuations of
various observables, some of the interesting ones have been discussed here.
The extraction of dynamical fluctuations originating from QGP
phase transition from the experimental results
becomes complicated because of several competing processes. This has been
addressed from the available data on 
particle multiplicity, net charge, strangeness, and  $\ave{p_{\rm T}}$.
Differential measures are being adopted in order to gain insight to
the details of fluctuation. Analysis based on forward-backward
long range multiplicity correlations have recently been performed 
\cite{EBYE:phobos,EBYE:brijesh} which show the presence of
significant correlations for central collisions.

One of the most important aspects of QGP study is the location of the critical point.
It may be possible to access this experimentally by scanning the QCD phase diagram
in terms of baryon chemical potential and temperature. This can be accomplished 
by varying beam energies 
from about $\sqrt{s_{\rm NN}}$=5~GeV to 100~GeV. 
Such a program is recently being undertaken
at RHIC \cite{EBYE:rhic}. Experiments at GSI \cite{EBYE:CBM} are planned to study this as well.
At higher energies of LHC
(\mbox{Pb--Pb} beams at $\sqrt{s_{\rm NN}}$=5.5~TeV), the ALICE experiment 
will be able to make
precise event-by-event measurements of various quantities and study their fluctuations 
\cite{EBYE:alice_ppr}. With continued development in new analysis methods and
theoretical advances, and with dedicated experiments, one will certainly
learn a great deal more about QGP phase transition through 
fluctuation studies.


\section*{References}
\begin{thebibliography}{10}

\bibitem{EBYE:rev}
    H. Heiselberg, Physics Reports {\bf 351} (2001) 161.
\bibitem{EBYE:tricrit}
  M. A. Stephanov, K. Rajagopal and E. Shuryak, Phys. Rev. Lett. {\bf 81} (1998) 4816.
\bibitem{EBYE:wa98_ebye}
    M.M. Aggarwal {\it et al.}, (WA98 Collaboration), Phys. Rev. {\bf C65}, (2002) 054912.
\bibitem{EBYE:na49_mult_fluc}
     M. Rybczynski {\it et al.} (NA49 Collaboration), J. Phys. Conf. Ser. {\bf 4}, (2005) 74.
\bibitem{EBYE:phenix_mult_fluc}
     J. Mitchell {\it et al.} (PHENIX Collaboration), {\it Preprint} nucl-ex/0510076.
\bibitem{EBYE:As00} 
    M. Asakawa, U. Heinz, B.Muller, Phys. Rev. Lett. {\bf 85} (2000) 2072.
\bibitem{EBYE:Ko00} 
    S. Jeon and V. Koch, Phys. Rev. Lett. {\bf 85} (2000) 2076.
\bibitem{EBYE:Sh01}
    E. Shuryak and M.A. Stephanov, Phys. Rev. {\bf C63} (2001) 064903. 
\bibitem{EBYE:evolution}
    B. Mohanty, J. Alam, T.K. Nayak, Phys. Rev. {\bf C67} (2003) 024904.
\bibitem{EBYE:na49_net} 
      C.~Alt {\it et al.}  (NA49 Collaboration), Phys. Rev. C\ {\bf 70} (2004) 064903.
\bibitem{EBYE:star_mult} 
         J. Adams {\it et al.}  (STAR Collaboration), Phys. Rev. {\bf C68} (2003) 044905.
\bibitem{EBYE:phenix_net} 
      K.~Adcox {\it et al.}  (PHENIX Collaboration), Phys. Rev. Lett.  {\bf 89} (2002) 082301.
\bibitem{EBYE:volo}
     C. Pruneau, S. Gavin, S. Voloshin, Phys. Rev. {\bf C66} (2002) 044904.
\bibitem{EBYE:mitchell}
     J.T. Mitchell, J. Phys {\bf G30} (2004) S819.
\bibitem{EBYE:ekr}
   S.\ Ejiri, F.\ Karsch and K.\ Redlich, Phys. Lett. B633 (2006) 275.
\bibitem{EBYE:ggnls}
   R.\ V.\ Gavai and S.\ Gupta, Phys. Rev.  {\bf D72} (2005) 054006.
\bibitem{EBYE:qm04k2pi}    
                  C. Roland {\it et al.} (NA49 Collaboration), J. Phys. {\bf G30} (2004) S1381.
\bibitem{EBYE:star_kpi} 
     S. Das {\it et al.} (STAR Collaboration), {\it Preprint} nucl-ex/0503023.
\bibitem{EBYE:torr}
     G. Torrieri, S. Jeon and J. Rafelski, {\it Preprint} nucl-th/0510024
\bibitem{EBYE:ceres_pt}
     Hiroyuki Sato {\it et al.} (CERES Collaboration), J. Phys. {\bf G30} (2004) S1371.
\bibitem{EBYE:na49_pt} T. Anticic {\it et al.} (NA49 Collaboration),
       Phys. Rev. {\bf C70} (2004) 034902
\bibitem{EBYE:phenix_pt}
     S.S. Adler {\it et al.} (PHENIX Collaboration), Phys. Rev. Lett. {\bf 93} (2004) 092301. 
\bibitem{EBYE:star_pt} J. Adams {\it et al.} (STAR Collaboration) Phys. Rev. {\bf C72} (2005) 044902.
\bibitem{EBYE:star_pt_etaphi} J. Adams {\it et al.} (STAR Collaboration) J. Phys. {\bf G32} (2006) L37.
\bibitem{EBYE:Pratt} 
      S. A. Bass, P. Danielewicz and S. Pratt, Phys. Rev. Lett. {\bf 85} (2000) 2689.
\bibitem{EBYE:star_bf} 
         J. Adams {\it et al.}, (STAR Collaboration) Phys. Rev. Lett. {\bf 90} (2003) 172301.
\bibitem{EBYE:na49_bf} 
         C. Alt {\it et al.} (NA49 Collaboration), Phys. Rev. {\bf C71} (2005) 034903;
\bibitem{EBYE:na49panos}
         P. Christakoglou {\it et al.} (NA49 Collaboration), {\it Preprint} nucl-ex/0510045.
\bibitem{EBYE:phobos}
         B.B. Back {\it et al.} (PHOBOS Collaboration) Phys. Rev. {\bf C74} (2006) 011901.
\bibitem{EBYE:brijesh}
        Terence Tarnowsky {\it et al.} (STAR Collaboration), {\it Preprint} nucl-ex/0606018
\bibitem{EBYE:rhic}
        T. Ludlum {\it et al.} BNL-75692-2006, 
        Proceedings of the workshop on 
        ``Can we discover the QCD Critical point at RHIC?'', March 9-10, 2006.
\bibitem{EBYE:CBM} 
        C. Hohne  {\it et al.} (CBM COllaboration) Nucl. Phys. {\bf A749} (2005) 141.
\bibitem{EBYE:alice_ppr} Physics Performance Report, Volume II, ALICE Collaboration, 
        to be published in J. Phys. G.
\end{thebibliography}
\end{document}